\documentclass[12pt]{article}
\usepackage{pdproc,epsfig,graphicx,psfig} 
\begin{document}

\renewcommand{\thefootnote}{\alph{footnote}}
  
\title{PUZZLES IN ASTROPHYSICS IN THE PAST AND PRESENT \footnote{
I dedicate this talk to D.V.Sciama, great personality and outstanding 
physicist}}

\author{ V. Berezinsky}

\address{INFN, Laboratori Nazionali del Gran Sasso,
Assergi (AQ) 67010 Italy\\ 
and Institute for Nuclear Research, Moscow\\
{\rm E-mail: berezinsky@lngs.infn.it}}

\abstract{About  400 years have passed since the great discoveries by
Galilei, Kepler and Newton,
but astronomy still remains an important source of discoveries in physics. 
They start with puzzles, with phenomena difficult to explain, 
and which in fact need for explanation the new physics. Are such puzzles
existing now? There are at least three candidates: absence of absorption of TeV
gamma radiation in extragalactic space (violation of Lorentz invariance?),  
absence of GZK cutoff in the spectrum of Ultra High Energy Cosmic Rays 
(new particle physics?), tremendous energy (up to $10^{54}$~ergs) released 
in Gamma Ray Bursts during a time scale of a second (collapsing stars 
or sources of a new type?). Do these puzzles really exist?  
A critical review of these phenomena is given.  }
   
\normalsize\baselineskip=15pt

\section{Introduction}
\begin{center}
        ALL GREAT DISCOVERIES IN ASTROPHYSICS\\
        APPEARED UNPREDICTABLY; WHAT WAS PREDICTED\\ 
        WAS NOT DISCOVERED.
\end{center}
\vspace{5mm}
Not many good things fall down on us from the sky, but discoveries do.
I will list below a short list of astrophysical discoveries of the last 
four decades, separating intuitively astrophysics from cosmology.\\*[3mm] 
{\it Quasars} were discovered in early 1960s as compact radio sources. 
Mathews and Sandage in 1960 identified radio source 3C48 with a
stellar-like object. Schmidt in 1963 deciphered the optical spectrum of
quasar 3C273 assuming its redshift, $z=0.158$. Surmounting resistance of 
sceptics, this explanation moved the source to
the distance of 630~Mpc and made its luminosity uncomfortably large,
$L \sim 10^{46}$~erg/s. This puzzling energy release resulted in the
long run in the discovery of a {\it black hole}, an object of
general relativity.  \\*[3mm]
{\it Pulsars} were discovered first in 1967 by a student of A. Hewish, 
Jocelyn Bell. She observed a puzzling periodicity of radiopulses from 
an unknown source. After short but intense discussion of different possible
sources, including extraterrestrial civilisations and ``little 
green men'', the magnetised
rotating neutron stars, the pulsars, were found to be responsible. It opened 
a new field of cosmic physics: {\em relativistic electrodynamics}.\\*[3mm] 
{\it The atmospheric neutrino anomaly and the solar neutrino problem} 
went along most difficult road  
to the status of discovery. The puzzling phenomenon in 
both cases was a  neutrino deficit as compared with calculations. But 
scepticism of the community, especially in the case  of the solar neutrino
problem  was strong. Pushed mostly by Davis and Bahcall, the solar
neutrino problem moved like  a slow coach along a road two decades
long. Fortunately, physics differs from democracy: opinion of
majority means usually less than that of ONE. These two obscure puzzles have
turned (or have almost turned) 
into discovery of the most fascinating phenomenon, {\it neutrino oscillations}.   
\\*[3mm]
{\it Supernova SN 1987a} became elementary-particle laboratory in the sky
for a study of 
properties of neutrinos, axions, majorons {\it etc}. 
Detection of neutrinos \cite{SN1987} became a triumph of the theory:
the number of detected neutrinos, duration of the neutrino pulse and 
estimated neutrino luminosity have appeared in agreement with
theoretical prediction. Gravitational collapse as a phenomenon
providing the SN explosion was confirmed. 

However, some puzzles remain. Presupernova was the blue supergiant, not the red
one as a theory of stellar evolution prescribes. But what is more puzzling
is rotation. The asymmetric ring around SN 1987a implies that the
presupernova was a rotating star (it would be a surprise if not!).     
But the striking agreement of neutrino observations with calculations
were obtained for a non-rotating presupernova. Inclusion of rotation 
in calculations is a very difficult task. The simplified calculations
\cite{ImNa} demonstrate that rotation changes predictions
dramatically: temperature of neutrinosphere diminishes by factor 2, 
total energy of emitted neutrinos becomes 6 times less and the number
of detected neutrinos should be an order of magnitude less.\\*[3mm]  
{\bf 1.1 Greatness of false discoveries}\\*[2mm]
\begin{center}
WHEN FIRST APPEARED THE PUZZLES LOOK WEAK.\\
SAVE YOUR TIME AND SAY: IT'S RUBBISH.\\
IN 90$\%$ OF CASES YOU WILL BE RIGHT. 
\end{center}
\vspace{3mm}
False discoveries often have greater impact on physics than the
true ones. 

In the end of 60s and the beginning of 70s using long baseline interferometry,
it was found that gas clouds in quasars and radiogalaxies in some 
cases had velocities exceeding the light speed by factor 4 - 10. 
In fact the measured velocity was a projection of velocity on the plane 
perpendicular to the line of view. Accurately written in relativistic 
mechanics, this (apparent) velocity is
\begin{equation}
v_{app}=\frac{v\sin\theta}{1-\frac{v}{c}\cos\theta}.
\end{equation}
Provided by ultrarelativistic velocity of an object $v\sim c$, the apparent 
velocity can exceed speed of light. {\it Astrophysics of relativistic
objects}, now a subject of university courses, was born. \\*[3mm] 
{\it Cyg X-3 saga} is a story of a different kind. 

Cyg X-3 is a galactic binary system well studied in all types of radiations,
most notably in X-rays. In 80s many EAS arrays detected from it 
4.8 hour periodic ``gamma-ray'' signal in VHE (Very High Energy, 
$E\geq 1$~TeV) and UHE (Ultra High Energy, $E\geq 0.1-1$~PeV) 
ranges. The list 
of these arrays included Kiel, Haverah Park, Fly's Eye, Akeno, 
Carpet-Baksan, Tien-Shan, Platey Rosa, Durham, Ooty, Ohya, Gulmarg, 
Crimea, Dugway, Whipple and others. Probably it is easy to say that 
there was no
single EAS array which claimed no-signal observation. Additionally, some
underground detectors (NUSEX, Soudan, MUTRON) marginally observed 
high energy muon signal from the direction of
this source. Apart from the Kiel array, which claimed $6\sigma$ signal,
the confidence level of detection was not high: $3-4 \sigma$.   

In 1990 - 1991 two new generation
detectors, CASA-MIA and SYGNUS, put the stringent upper limit to the signal 
from Cyg X-3, which excluded early observations. 

Apart from two lessons:\\
{\it (i)} good detectors are better than bad ones,\\
{\it (ii)} ``$3\sigma$'' discoveries should not be trusted, even if many 
detectors confirm them, \\ 
experience of Cyg X-3 has taught us how to evaluate statistical significance
searching for periodic signals. 

The false discovery of high energy radiation from Cyg X-3 had great
impact on theoretical high energy astrophysics, stimulating study of 
acceleration in binary systems, production of high energy gamma and neutrino
radiation  and creation of high energy astrophysics with new
particles, such as light neutralinos, gluinos {\it etc}.  

\section{Violation of Lorentz Invariance}

Violation of Lorentz invariance (LI) is often suspected in
astrophysics because of large Lorentz factors being sometimes involved. 
Before describing these suspicions, we will discuss the 
aesthetic side of the problem: is there an aesthetically attractive   
theory with a broken LI?  

Breaking of LI, even extremely weakly, leads to 
existence of the absolute Lorentz frame. This is a qualitative
difference between the two theories. Absence of continuous transition
from one theory to another looks disturbing.   

Lorentz invariance is a basic principle for building a Lagrangian for
any interaction. How is it possible to abandon it?

All questions raised above disappear in spontaneously broken LI. 
Equations of motion remain Lorentz invariant. The violation occurs 
spontaneously in the solutions. Lagrangians for all interactions are
constructed as Lorentz scalars and spontaneous LI breaking occurs due 
to non-zero values of field components in vacuum states. Breaking of 
LI can be made arbitrarily small, and all physical effects accompanied 
by LI breaking are small too. The absolute Lorentz frame exists, but 
all physical effects, which distinguish it from other frames, are
small, and thus all frames are nearly equivalent, similar to Lorentz 
invariant theory.\\*[3mm]
{\bf 2.1 Spontaneously broken Lorentz invariance}\\*[2mm]
The Lorentz invariance is spontaneously broken when  time component
of vector or tensor field obtains non-zero value. The necessary 
condition for the phase transition to such configuration is existence
of potential minimum at this value. Such a condition can be
fulfilled only in some exceptional cases, e.g in superstring theories 
\cite{string} and in some specific D-brane models with extra dimensions
\cite{Dvali,Ellis}. The interactions responsible for such potential minimum 
usually do not appear in conventional four-dimensional renormalizable
theories.   

Consider for example the Lorentz invariant interaction of superheavy 
tensor field $T_{\mu\nu...}$ with ordinary field described by spinor 
$\psi$. If, for example, string interactions set non-zero vev for time
components of this tensor field, the considered interaction term is 
reduced to the term which explicitly breaks LI \cite{CoKo97}:
\begin{equation}  
{\cal L}_{int}=\frac{\epsilon}{M^k}v_k \bar{\psi}\Gamma (i\partial_0)^k\psi,
\label{LIbr}
\end{equation}
where $v_k=<T_{00...}>$ is vev, $\Gamma$ is build from $\gamma$'s ,
$\epsilon$ is dimensionless constant, and $M$ is a superstring mass
scale. Such LI breaking term modifies dispersion relation for 
a particle $\psi$ and results in astrophysical consequences
\cite{CoGl}.

We shall give here a simple example of spontaneous LI breaking accompanied
by modification of dispersion relation for an ordinary particle. 

Let us consider the Lagrangian for an ordinary spinor particle
\begin{equation}
{\cal L}=i\bar{\psi}\gamma_{\mu}\partial_{\mu}\psi 
-m\bar{\psi}\psi +{\cal L}_{int},
\label{Lagr}
\end{equation}
where interaction with superheavy field $T_{\mu\nu}$ is described by 
\begin{equation}
{\cal L}_{int}=\frac{\epsilon}{M^2}T_{\mu\nu}\bar{\psi}\gamma_{\mu}
\partial_{\nu}\psi.
\label{inter}
\end{equation}

After spontaneous symmetry breaking $<T_{00}>=v^2$
the associate Klein-Gordon equation \cite{KG} for a non-interacting
particle can be readily obtained as
\begin{equation}
(\partial_{\mu}^2 +m^2- \frac{\epsilon}{M^2}v^2E^2)\psi=0
\end{equation}
The corresponding dispersion relation is 
\begin{equation}
p_{\mu}^2-m^2+\frac{\epsilon}{M^2}v^2 E^2=0.
\label{DR}  
\end{equation}
{\bf 2.2 Modified dispersion relations and threshold of reactions}\\*[2mm]
In astrophysical applications one often considers a collision of
very high energy particle with low energy particle from background
radiation. This is the case of  absorption of TeV
gamma-radiation on infrared photons and so-called GZK\cite{GZK} cutoff, when
very high energy proton collides with a microwave photon, producing the
pion. A threshold of such reactions is determined by momentum of a  particle
in c-m system, which has 
very large Lorentz factor in laboratory system. At large Lorentz
factors, {\it i.e.}
at large energy of one of colliding particles in laboratory system, the
dispersion relation is modified and affects the threshold of reaction
in laboratory system.

Let us consider this effect in $\gamma+\gamma \to e^+ +e^-$ collision 
when one photon has large energy $E_{\gamma}$ and another - small 
energy $\epsilon_{\gamma}$. The conservation of energy and momentum 
requires 
\begin{equation} 
(k_{\mu}+k_{\mu}')^2= (p_{\mu}+p_{\mu}')^2,
\label{conserv}
\end{equation}
where $k_{\mu}$ and $p_{\mu}$ are 4-momenta of photon and electron, 
respectively, and their 4-momenta after collision are shown by primes. 
In LI case $k_{\mu}^2=k_{\mu}^{'2}=0$ and $p_{\mu}^2=p_{\mu}^{'2}=m_e^2$, 
and energy-momentum conservation requires
\begin{equation}
\sqrt{E_{\gamma}\epsilon_{\gamma}}> m_e
\label{thresh}
\end{equation}
as the threshold condition. In the case of modified dispersion relation 
$p_{\mu}^2=E_e^2-\vec{p}_e^2\approx m_e^2+ \epsilon v^2 E^2/M^2$, 
as in Eq.(\ref{DR}), and the threshold shifts towards higher energies
in laboratory system.\\*[2mm]
{\bf 2.3 Astrophysical tests of special relativity}\\*[2mm]
GZK cutoff involves Lorentz transformations with Lorentz factor 
$\Gamma \sim m_{\pi}/\epsilon_{\gamma} \sim 10^{11}$, where $m_{\pi}$ is 
a pion mass and $\epsilon_{\gamma} \sim 10^{-3}$~eV is a typical
energy of microwave photon participating in the photopion
reaction.  It gives the largest Lorentz factor presently known, up to
which LI can be tested. Such a test and proposal to explain the
absence of GZK cutoff in experimental data were first suggested by 
Kirzhnitz and Chechin \cite{Kirzh} in 1972 and later in 
Refs.\cite{CoGl,Gonz,Glash,Grillo}.

The other similar process is absorption of TeV gamma radiation on 
IR background photons. It will be considered at some details in the
next Section.

There are some other astrophysical processes, where LI can reveal
itself:
\begin{itemize}
\item Constancy velocity of light \cite{ElNa} 
\item Vacuum Cherencov radiation \cite{Glash},
\item Vacuum Faraday rotation \cite{Glash},
\item High energy photon decay \cite{Glash},
\end{itemize}

\section{TeV Gamma-ray Crisis?}

Propagating through the space filled by IR background radiation, TeV
photons are absorbed in $\gamma\gamma_{IR} \to e^++e^-$ collisions. 
A photon with energy $E_{\gamma}$ is absorbed by IR photons with wavelengths  
shorter than 
\begin{equation}
\lambda_{IR} \approx \frac{E_{\gamma}}{4m_e^2} = 
1.2 \frac{E_{\gamma}}{1~{\rm TeV}}~\mu{\rm m}.
\end{equation}
Thus photons with energies 1 - 20~TeV are absorbed on IR background
radiation with wavelengths 
$1\mu{\rm m} \leq \lambda_{IR} \leq 20 \mu{\rm m}$. For this  
wavelength range there are both direct measurements and detailed
calculations. 
  
\begin{figure}[htb]
\begin{center}
\centering
\psfig{bbllx= 70pt, bblly=540pt, bburx=730pt, bbury=60pt,
file=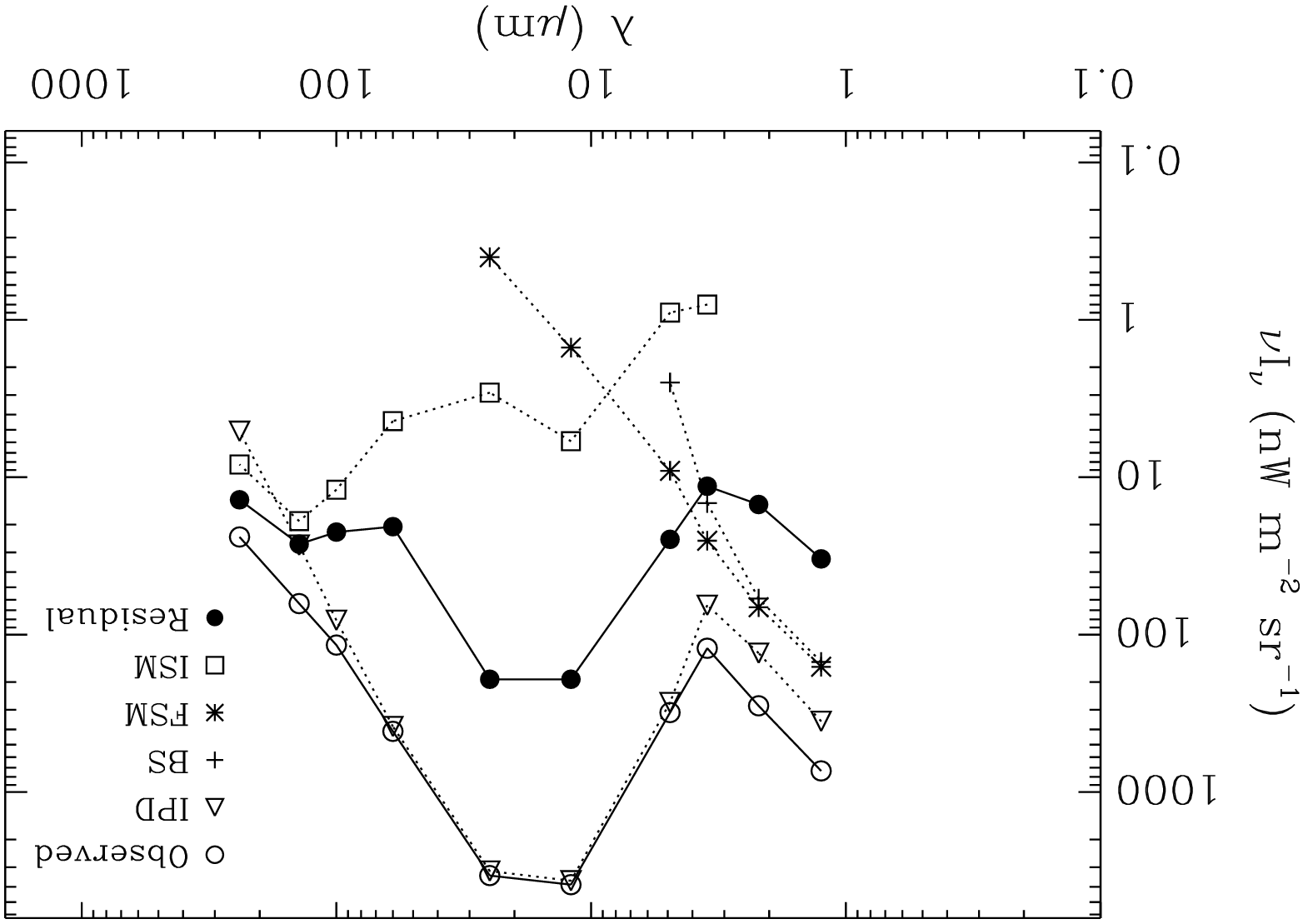, height=8.0cm ,angle=90}
\end{center}
\vspace{80mm}
\caption{\em IR diffuse spectra as measured by
DIRBE-COBE in the range 1.25 - 240~$\mu$m (from \protect\cite{DIRBE}). 
Extragalactic diffuse flux, shown by filled circles, is
obtained by subtraction of contributions due to Interplanetary dust (IPD)
shown by triangles, bright galactic sources (BS),  shown by crosses, and
others.}
\label{dirbe}
\end{figure}

The DIRBE instrument\cite{DIRBE} (Diffuse Infrared Background Experiment) on 
COBE spacecraft measures IR diffuse radiation in the band from 
1.25 to 240~$\mu{\rm m}$. The measured fluxes are shown in Fig.\ref{dirbe}. 

\vspace*{10mm}
Note that the measured flux is very close to that produced by
interplanetary dust and after subtraction of this major component the
residual is about factor of 20 less. 

The data of another COBE instrument, FIRAS (Far Infrared Absolute
Spectrometer) in the range 125 - 2000~$\mu$m  are consistent \cite{FIRAS} 
with DIRBE at 
overlapping frequencies. An estimate of intergalactic diffuse IR  
flux from integrating the 15 microns count of IR sources at 
ISOCAM \cite{ISOCAM} is also consistent with the DIRBE data. 
However, a count of sources is always incomplete, and thus the ISOCAM
flux at 15~$\mu$m can be considered as an lower limit. 

\begin{figure}[htb]
\epsfxsize=11truecm
\centerline{\epsffile{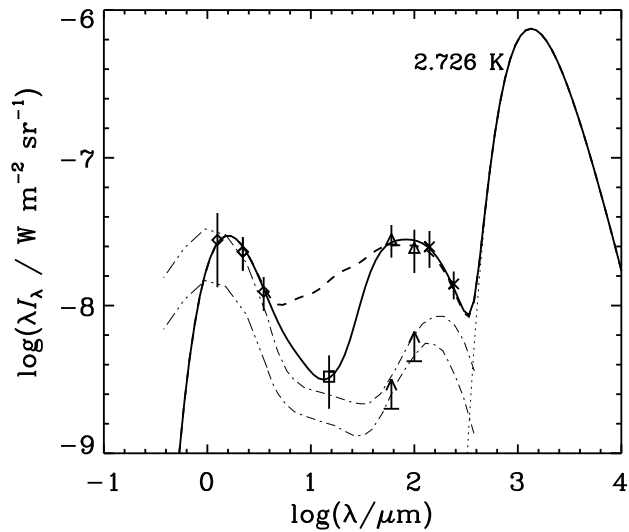}}   
\vspace{-5mm}
\caption{\em Intergalactic diffuse IR fluxes \protect\cite{Proth}. 
The data points are from
DIRBE (diamonds), ISOCAM (square) and FIRAS (crosses). These data can
be described by a thick solid line or by a solid line with a broken
line in the middle. Two theoretical curves (dot-dash ones) have emerged from
calculations by Malkan and Stecker \protect\cite{Steck}. 
}
\label{IR-proth}
\end{figure}

Diffuse IR flux was calculated in many works, most notably in two 
recent works \cite{Steck,Prim} (see the references to early
calculations there). In Ref.\cite{Steck} a semi-empirical method is
used, while calculations of Ref.\cite{Prim} are based on galaxy formation
models. A feature common to both calculations is bimodal frequency
distribution. A peak at about 1~$\mu$m is a direct radiation during
star formation epoch, and a second peak about 100~$\mu$m is due to
dust re-radiating the starlight at lower frequencies. It could be 
(see Fig.\ref{IR-proth}) that observational data confirm this
feature.   
Both calculations give IR flux lower than that of DIRBE. 

The results of observations and comparison with
calculations are given in Fig.\ref{IR-proth}, taken from Ref. \cite{Proth}.
\vspace{-5mm}  

\begin{figure}[htb]
\epsfxsize=11truecm
\centerline{\epsffile{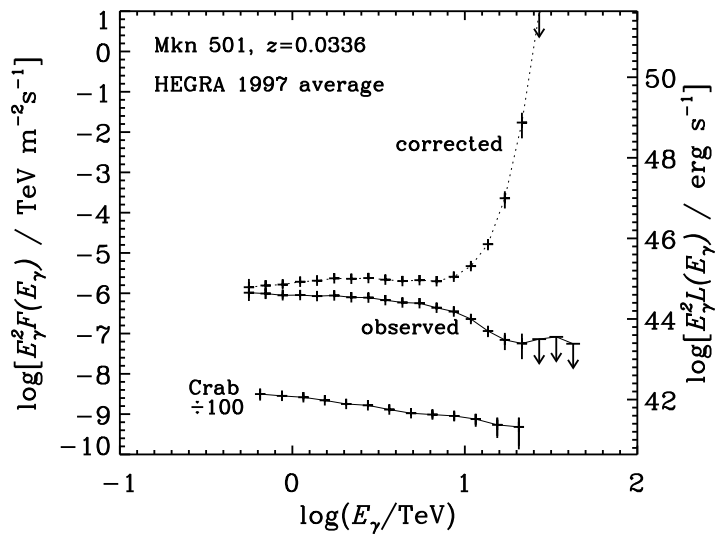}}   
\caption{\em The production (``corrected'') spectrum for Mrk 501 calculated 
\protect\cite{Proth} from the observed spectrum and with $\gamma\gamma$
absorption taken into account. Unnaturalness of production spectrum is
illustrated by luminosity shown on the right-hand axis. }
\label{Tev-proth}
\end{figure}

As was discussed above, the gamma-radiation with energy 1- 10 ~TeV is
absorbed by IR radiation in the range 1 - 10~$\mu$m. A gamma-ray
source with  relevant properties is a nearby blazar Mrk 501. This 
powerful TeV source is located at suitable distance 155~Mpc (z=0.0336) and its
measured spectrum extends from 400~GeV up to 24~TeV without noticeable
steepening. However, the expected absorption in the end
of the spectrum is appreciable, and to have the observed spectrum
without cutoff, the production (``corrected'') spectrum must be
unnatural, as it is shown in Fig.\ref{Tev-proth} from Ref. \cite{Proth}.
In fact, this unnatural increase in production spectrum corresponds 
to spectrum cutoff, if the production spectrum is smooth. The cutoff
appears  when a pathlength diminishes with energy, as it happens when 
density of target photons increases with wavelength $\lambda$. This is 
the case of dependence of IR flux on $\lambda$ presented in 
Fig.\ref{IR-proth}. 

\begin{figure}[htb]
\begin{center}
\psfig{bbllx= 70pt, bblly=230pt, bburx=470pt, bbury=600pt,
file=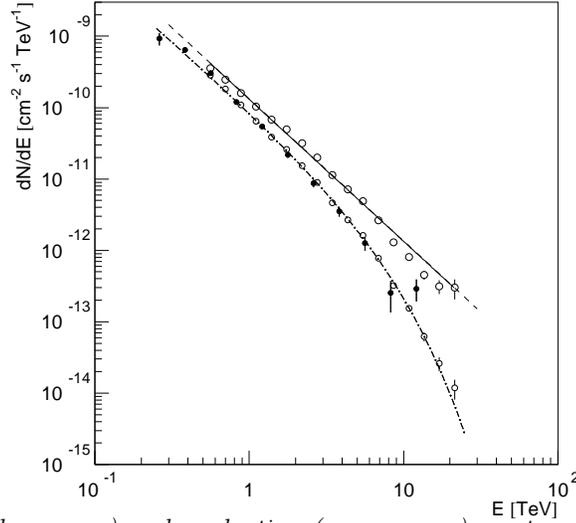, height=8.0cm ,angle=0}
\end{center}
\vspace{-10mm}
\caption{\em The observed (low curve) and production (upper curve) 
spectra of TeV gamma radiation for Mrk 501 (from \protect\cite{Steck}). 
The production spectrum is
calculated using the absorption on IR radiation with the flux
calculated in one of the models of Ref.\protect\cite{Steck}.  
}
\label{steck}
\end{figure}

If IR flux is taken according to theoretical calculations \cite{Steck,Prim}, 
``TeV gamma-ray crisis'' disappears. In Fig.\ref{steck} the observed
spectrum Mrk 501 in the flaring state (low curve) is given together
with the production spectrum (upper curve). The absorption in 
intergalactic space is taken according to IR flux calculated by the authors 
of Ref.\cite{Steck} in one of their models. The source spectrum is quite
natural with no indication to the ``crisis''. The similar calculations
were made in Ref.\cite{Ahar}. The source (production) spectrum of 
TeV radiation from Mrk 501 was calculated from the observed spectrum 
using the theoretical IR flux, according to LCDM model of
Ref.\cite{Prim}. The production spectrum is found to be natural. 
\vspace{-5mm}
\vspace{5mm}    

In conclusion, ``TeV gamma-ray crisis'' does not look dramatic. It 
could be that IR flux measured by DIRBE is slightly overestimated due
to incomplete subtraction of galactic or interplanetary components.  
The frequency dependence of the flux plays crucial role. The spectrum
cutoff appears if flux increases with $\lambda$ sharply enough. If ISOCAM  
point in Fig.\ref{IR-proth} is in fact an lower limit, the frequency
dependence of IR flux can be smooth (as the broken line shows), and the
sharp cutoff is absent. 

At present there is no need to involve Lorentz Invariance breaking
for solving this problem. A clear test of Lorentz invariance can be
done with help of radiation for which the density of target photons 
and their energy spectrum are reliably known. This case is given by 
microwave relic radiation. As Fig.\ref{IR-proth} shows at wavelength 
$\lambda \sim 300~\mu$m there is a sharp increase of photon density
with $\lambda$, which results in the cutoff of gamma-ray spectrum at 
$E_{\gamma} \sim 250$~TeV. An absence of such cutoff for a source with 
known distance to our galaxy (redshift) can be explained only by
Lorentz Invariance breaking. 
\newpage
\section{UHECR: How Serious is the Problem?}
The problem with Ultra High Energy Cosmic Rays (UHECR) consists in 
observation of particles with very high energies, up to 
$2-3\cdot 10^{20}$~eV, while the ordinary signal carriers such as
protons, nuclei, electrons and photons have a small pathlength in intergalactic
space.  UHE protons loose energy due to production of pions in
collisions with microwave photons and their spectra should have a steepening
(GZK cutoff \cite{GZK}) which starts at $E \sim 3\cdot 10^{19}$~eV. 
UHE nuclei loose energy due to $e^+e^-$ pair production in collisions 
with microwave photons \cite{BGZ}. Electrons are loosing the  energy
very fast on microwave radiation. UHE photons are absorbed on extragalactic 
radio background \cite{Be1970}. 

UHE particles are observed by Extensive Air Showers (EAS) produced in the
atmosphere. 

Doubts in existence of UHECR problem are usually expressed in the form
of two questions:
\begin{itemize} 
\item  Are energies measured correctly?
\item  Could the sources of UHECR be located nearby, e.g. in our
galaxy or at small distance from it?
\end{itemize}

I will address below these questions and analyse the status of UHECR problem.
\\*[2mm]
\noindent
{\bf 4.1 Energy determination}
\\*[2mm]
The energy of a primary particle is determined measuring some characteristics
of EAS (for a review see \cite{NaWa}). The methods
are different and agree for determination of UHE energies within $20-30\%$.
The error in energy determination is estimated as $15-20\%$ for the good 
events. 
The most traditional method of energy measurement is based on the 
relation between cascade
particle density at the distance 600 m from the shower axis  and  primary
energy E. This method ($\rho_{600}$) was first 
suggested by Hillas \cite{Hillas}, and 
later confirmed my many Monte Carlo simulations. It was demonstrated that
this relation depends weakly on the model of EAS development and on 
chemical composition of the primaries. The density fluctuations have
also minimum at the distance 600 m from the core.  For UHE EAS 
this relation between $\rho_{600}$ and primary energy has been  
confirmed by calorimetric measurements at Yakutsk for 
energies up to $4\cdot 10^{18}$~eV. The $\rho_{600}$ method was used 
in the Haverah Park, Yakutsk and AGASA arrays. In the case of Haverah
Park $\rho_{600}$ signal is given by energy release in water
Cherenkov detectors. 

The Fly's Eye array detects fluorescence light produced by EAS in the
atmosphere. The intensity and arrival time of fluorescent
radiation to the collecting mirrors allow to reconstruct the longitudinal 
development of EAS in the atmosphere, and the primary energy is
obtained thus practically calorimetrically. 

In two cases the primary energy was measured very reliably. 

The Fly's Eye detector had an event \cite{FE} with very accurately measured 
longitudinal profile and the primary energy was found to be 
$E=(3^{+0.36}_{-0.54})\times 10^{20}$~eV. This is the highest energy 
event. 

The AGASA array detected \cite{AGASA} a shower with the core in the
dense part of
array, Akeno. The lateral distribution of cascade particles, including
muons, was measured in the total range of distances from the core up to 
3~km. The primary energy is estimated to be in the range 
$(1.7 -2.6)\times 10^{20}$~eV.
 
The total number of detected showers with energy higher 
than $1\cdot 10^{20}$~eV is about 20. A sceptic taking event by event 
could doubt in energy of some of them, but even at most critical 
analysis several of them survive as cases with energy higher than 
$1\cdot 10^{20}$~eV. It is already enough to claim  existence of 
the problem. 
\\*[2mm]
{\bf 4.2 What is the GZK cutoff?}
\\*[2mm]
The Greisen-Zatsepin-Kuzmin cutoff \cite{GZK} is caused by interaction
of high energy protons with microwave (2.73 K) radiation. At energy 
$3\cdot 10^{19}$~eV the total energy loss of the proton starts 
sharply (exponentially) increasing with energy due to pion production 
$p+\gamma_{2.7K}\to N+\pi$. The exponential character of this increase
is caused by the fact that production of pions needs the photons from
high energy tail of their distribution, and the number of these
photons exponentially increase when proton energy becomes higher. 
The similar phenomenon occurs for nuclei Z, but the relevant process
is pair-production, $Z+\gamma_{2.7K}\to Z+e^++e^-$ \cite{BGZ}. 
\begin{figure}[htb]
\centering\includegraphics[height=8cm]{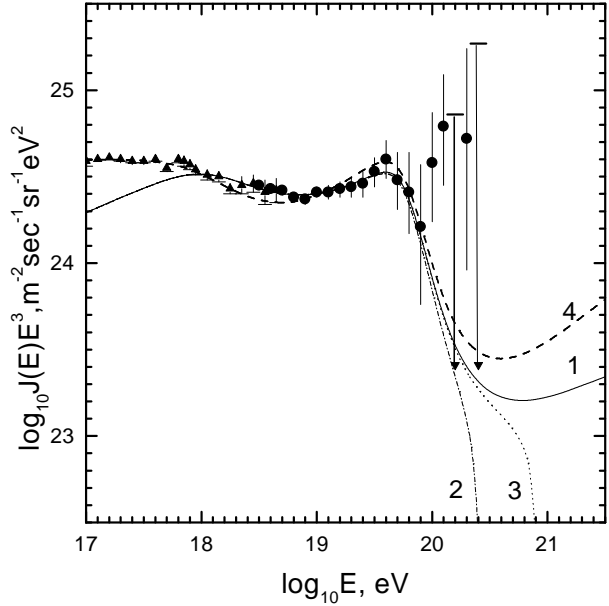}   
\caption{\em Calculated diffuse spectra \protect\cite{BGG} for 
uniform distribution of UHECR sources in the Universe in case of
cosmological evolution of the sources (curve 4) and without evolution 
(curves 1-3). The latter are given for different cutoff energies in 
the production spectra ($E_{\rm max}=\infty,~ 3\times 10^{20}$~eV
and $1\times 10^{21}$~eV, for curves 1,~2 and 3,respectively 
(see also \protect\cite{SS}).
}
\label{spectra}
\end{figure}
If a source is at the large distance this process causes the cutoff in
the observed spectrum. Numerically the energy of this cutoff is
usually given for a model where sources are distributed uniformly in 
extragalactic space . In this case the cutoff starts at 
$E=3\cdot 10^{19}~$eV and the flux becomes half of its power-law 
extrapolation at $E_{1/2}=5.3\cdot 10^{19}$~eV (for these and other
details of energy losses and absorption of different kind of primaries 
see Ref.\cite{book}). 

In Fig.\ref{spectra} the calculated spectra of protons are displayed 
for a model where the sources are distributed uniformly in the
Universe with cosmological evolution of the sources (curve 4) and
without it (curves 1 - 3). The evolutionary case is given for 
generation spectrum with spectral index $\gamma_g=2.45$ and with 
evolution described as $(1+z)^m$ with $m=4$. The case without evolution,
$m=0$, is presented for power-law generation spectrum with index 
$\gamma_g=2.7$ and 
with different cutoffs of generation spectra
described by maximal energies $E_{max}= 3\cdot 10^{20},~ 1\cdot 10^{21}$, and 
$ \infty$ \cite{BGG}.   The calculations are compared with 
recent AGASA data. The GZK cutoff in the calculated spectra is clearly
seen, in contrast to the observed spectrum. 
\\*[2mm]
{\bf 4.3 Extragalactic UHECR from Astrophysical Sources}
\\*[2mm]
An often asked question is: 
There could be a few extragalactic sources so close to us, that the observed
spectrum does not suffer the GZK cutoff, what is the problem then?

The problem is the low energy part of the spectrum. It is formed by 
sources at large distances, and because of GZK cutoff these sources 
do not contribute to high energy part of the observed
spectrum. One must specify his assumption about distribution of
sources in the universe, and the uniform distribution is the simplest
one.\\    
\begin{figure}[t]
\centering\includegraphics[height=8cm]{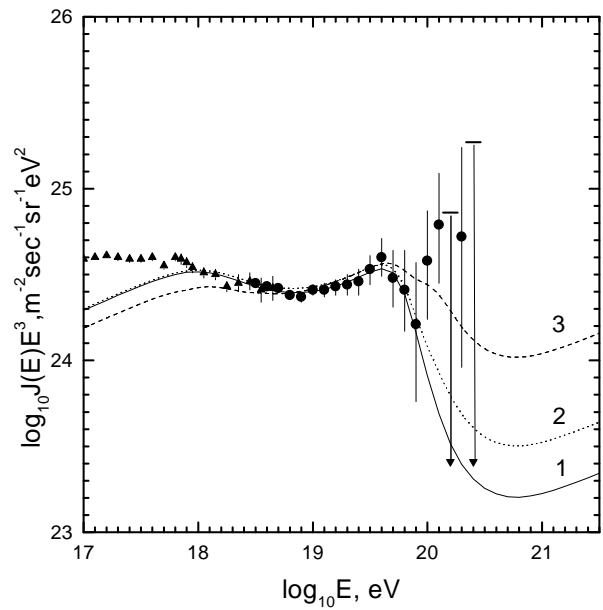}   
\caption{\em Diffuse energy spectra with local overdensity of UHECR
sources. Curves 1,~2 and 3 are given for uniform distribution of 
the sources, for overdensity $n/n_0=2$, and $n/n_0=10$, respectively. 
The linear size of overdensity region is 30~Mpc. 
}
\label{overd}
\end{figure}
GZK cutoff shifts to higher energies and becomes softer if population
of sources has local (within 30 - 50~Mpc) overdensity. 
(for early calculations see \cite{BGLS}, described in \cite{book}). The
count of galaxies show 
local overdensity with a factor $\sim 2$  within Local Supercluster.
{\em Since overdensity is a gravitational 
phenomenon, one must expect the similar overdensity for all galactic-like 
sources of UHECR , i.e. ones with  acceleration by stars, AGN and other
intragalactic objects.} In Fig.\ref{overd}
the spectra of UHECR are shown for three cases: uniform distribution
of the sources (curve 1), distribution with local overdensity 
$n/n_0 =2$ within radius $R=30$~kpc, and with local 
overdensity $n/n_0 =10$ within radius $R=30$~kpc. The
calculations of spectra for observed distribution of galaxies were
performed in Ref.\cite{BlOl}. From Fig.\ref{overd} one can
see that to reconcile the data with observations, the overdensity 
of the sources must be considerably larger than that  observed for 
the galaxies. This disfavours the idea.
Another possibility to make GZK cutoff less pronounced is given 
by one-source model.  

Let us assume that there are a few sources of UHECR in the Universe 
and by chance our Galaxy is located nearby one of them. In this case 
GZK cutoff is absent, but the price to be paid is anisotropy. In early  
work \cite{Wo} the Virgo cluster was considered as such source, and
energy of cutoff and anisotropy was reconciled assuming the diffusive
propagation of UHE particles. With new data on particle energies  and 
anisotropy this model is excluded. An interesting revival of this idea 
was suggested in Ref.\cite{Bier}. Particles generated in M87 galaxy in
Virgo cluster are propagating almost rectilinearly and are focusing to
the Sun by galactic magnetic field. It provides arrival to the Sun 
from different directions in agreement with absence of large
anisotropy. Quasi-rectilinear propagation from Virgo provides absence
of GZK cutoff.

A model of a nearby source with non-stationary diffusion was suggested
in Ref.\cite{BGD}
\\*[2mm]
{\bf 4.4 Galactic Origin of UHECR}
\\*[2mm]
UHECR have no GZK cutoff in case of galactic origin. Starting with 
S.I.Syrovatsky \cite{Sy}, the galactic origin of UHECR was advocated
in many works (see for example \cite{Chr,Gill,Pt}). Observational data 
are in favour of pure proton composition at highest energies.
\begin{figure}[h]
\centering\includegraphics[height=7cm]{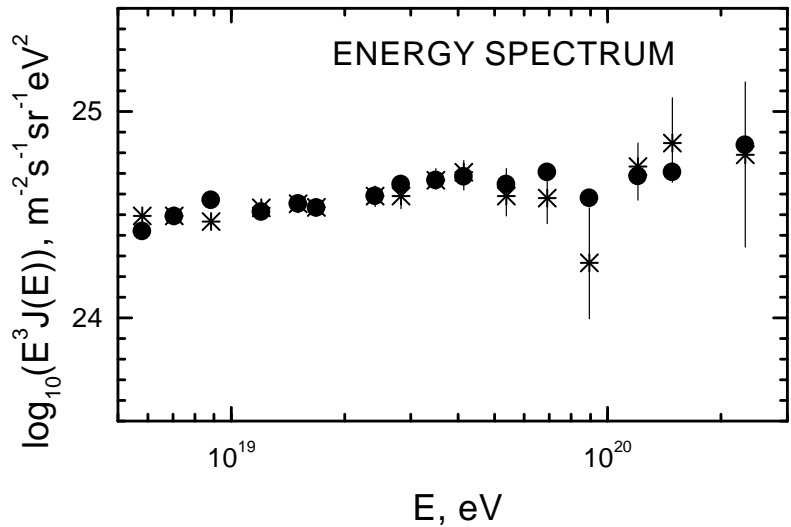}   
\caption{\it Energy spectrum of UHE iron nuclei of galactic origin
according to simulation \protect\cite{BGM} for generation spectrum
with $\gamma_g=2.3$. Black circles show the calculated flux 
(ad hoc normalisation), stars -- data of AGASA.
}
\label{gspectrum}
\end{figure}
As numerous simulations show, Galactic magnetic field cannot confine 
UHECR in the Galaxy, if they are protons. A case when they are 
heaviest (iron) nuclei is more difficult for a conclusion.
We shall present here results of the simulation of propagation of UHE
particles in our Galaxy \cite{BGM} with  conclusions concerning the iron
nuclei as primaries. 

The model is similar to one used in Ref.\cite{COSMAG}. The disc with 
UHECR sources is surrounded by  extended spherical halo with radius 
which varies from 15~kpc to 30~kpc. Magnetic field in the disc has 
a spiral structure with a spiral arms and is 
given by complicated analytic expression, which fits observational data.  
The field is dominated by the azimuthal component.
The thickness of magnetic disc is 0.4~kpc. The magnetic field in the halo 
is taken according to the theoretical model \cite{SS} and is described by
complicated analytic formulae (see also \cite{COSMAG}). The flux of
UHE particle from given direction is calculated in the following way. 
Antiparticle with energy $E$ is emitted in this direction and its
trajectory if followed step by step until exit from the halo. A
particle can cross the disc several times due to deflection by 
magnetic field in the halo. 
The intensity in given direction is proportional to the total time, $T_d$, 
a particle spends in the disc. The energy spectrum is given by the
product of generation spectrum $KE_g^{-\gamma_g}$ and $T_d(E)$. 
The calculated spectrum with $\gamma_g=2.3$ is shown in Fig.\ref{gspectrum} 
in comparison with AGASA data. One can see excellent agreement.

\begin{figure}[hbt]
\centering\includegraphics[width=10cm]{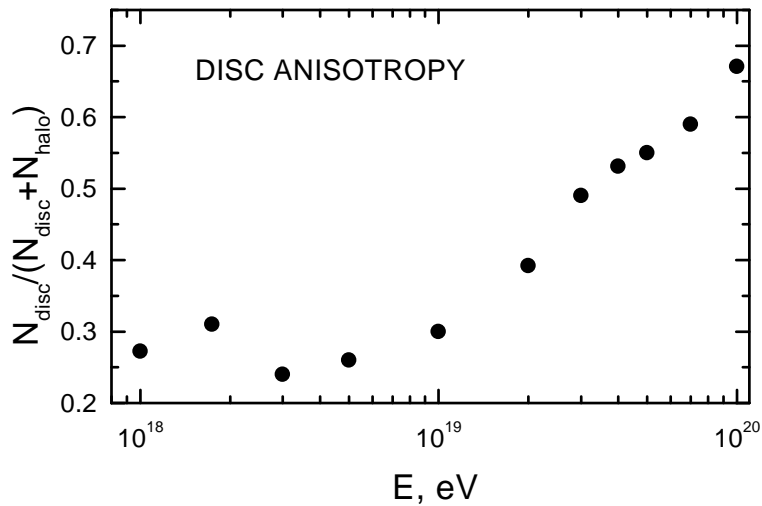}   
\caption{\em Disc anisotropy of UHE iron nuclei according to
simulation \protect\cite{BGM}}
\label{anisotropy}
\end{figure}

However, the real problem is given by anisotropy. In Fig. \ref{anisotropy}
the disc anisotropy is calculated as the ratio of the flux from
direction of the disc to the total flux. At energy 
$E\geq 4\cdot 10^{19}$~eV the calculated
anisotropy exceeds the observed value.\\
\\*[2mm]
{\bf 4.5 New physics}
\\*[2mm]
Although is not yet excluded, the astrophysical solution to UHECR problem 
is strongly disfavoured. We shall shortly describe here the 
elementary-particle solutions not giving the references (for a review 
of elementary-particle solutions see \cite{Be99}) 
\begin{itemize}
\item {\it Superheavy Dark Matter}. Long-lived Superheavy Dark Matter 
Particles are accumulated in galactic halos. These particles are 
naturally produced at post-inflationary epoch and can close the Universe or  
contribute some fraction to Cold Dark Matter. These particles can be 
long-lived, 
with lifetime exceeding the age of the Universe. Decays of these particles 
produce UHECR without GZK cutoff (most of UHE particles come from 
Galactic halo). The dominant component is photons.
\item {\it Topological Defects} (TD). There are different mechanisms of 
production of UHE particles by TD. In some cases TD become unstable 
and decompose to constituent fields (superheavy Higgs and gauge bosons),
which then decay to ordinary particles. This mechanism works for cusps 
and superconducting cosmic strings. In case of monopoles and 
antimonopoles connected by strings, high energy particles are produced 
at annihilation of monopole-antimonopole pairs. The most promising 
candidates are necklaces and  monopole-antimonopole pairs connected
by string. UHECR from TD has spectrum with a soft GZK cutoff which does 
not contradict observations.
\item {\it Resonant neutrinos}. Very high energy neutrinos are resonantly 
absorbed by target neutrinos comprising Hot Dark Matter (HDM): 
$\nu+{\bar \nu}_{HDM} \to Z^0 \to hadrons$. In the case HDM neutrinos have
locally enhanced density, the GZK cutoff is absent or softened. 
Very large flux of primary neutrinos with superhigh energies are
needed for this hypothesis.
\item {\it Light gluino}. Light gluinos can be effectively produced 
by TD or in pp-collisions in astrophysical sources. They weakly 
degrade in energy interacting with microwave radiation. The
interaction of UHE light gluino with nucleons is similar to that of
UHE proton. Light gluino is disfavoured by accelerator experiments. 
\item {\it Strongly interacting neutrino}. In extra-dimension
theories, for example,
neutrino can have large cross-section of scattering off the nucleon.
In this case neutrino can be a carrier of UHE signal from remote
astrophysical sources.
\item {\it Lorentz Invariance breaking}. In this case for protons with
energies $10^{20}$~eV and higher, the c.m. energy could be not enough
for production of pions in collisions with microwave photons (see
section 2).
\end{itemize}
\section{Mystery of GRB engine}
Gamma Ray Bursts (GRBs) strike our imagination. On one hand, the
tremendous energy release up to $2\times 10^{54}$~ergs during a short time
of order of a few seconds, within the volume of order of the moon,
implies a catastrophic event in the Universe. On the other hand we
observe them once a day. All observed characteristics of GRBs vary
within very wide range, fluences -- within $10^{-8} - 10^{-4}$erg/s,
durations -- from a few ms up to a few thousand of sec, energy
release -- from $10^{50}$ to $10^{54}$ ergs in case of isotropic radiation.  
It is very plausible that there are several types of GRBs of different 
origin. In particular, the short bursts, shorter than 1s, and long
bursts form two distinctive groups of GRBs.   

The mechanism of radiation in GRBs is well understood. The unknown
compact GRB engine explosively produces a fireball expanding in the 
surrounded plasma. An important parameter which determines the 
hydrodynamic expansion is the inverse baryon content, $\eta$, of 
the initial fireball, which is given by relation 
$\eta={\cal E}_{GRB}/M_bc^2$, 
where ${\cal E}_{GRB}$ is the total energy of GRB and $M_b$ is a baryonic
mass. Value of $\eta$ gives Lorentz factor of a fireball at the stage 
of saturation, which follows acceleration stage of expansion. 

Electrons, accelerated in the fireball by the
multiple internal shocks and as well as by the external and reversed shocks,
produce soft gamma-ray radiation by synchrotron mechanism. When 
the external shock starts decelerating, the synchrotron emission of 
electrons occurs in X-ray and optical range. This is so called afterglow 
radiation .

For most of GRBs the host galaxies are not found. It could be
explained by large  error box determined for GRBs and by weakness 
of host galaxies. Only in 1996 Beppo-SAX discovered the first host
galaxy due to afterglow radiation. Since that time for more than 20 GRBs the
host galaxies were reliably found. All galaxy-hosted GRBs have  
long durations and are located at large distances. 

The Beppo-SAX discovery gave  us a clear indication to the location of
the sources: at least some of them (and maybe all) are the galactic
objects. For astrophysical objects the greatest energy release is
given by gravitational collapse. In principle, energy release 
at the collapse can reach 29\% of the collapsed mass in case of a
black hole with maximal rotation. It gives $\sim 1\cdot 10^{54}$~erg  
for a collapse of $1~M_{\odot}$ stellar core, but it is difficult to
imagine that all this energy can be transferred to GRB. 

Strong beaming can solve the energy problem and this 
beaming looks like an inevitable element of any realistic
astrophysical model of GRB engine.

The moderate beaming is a common feature observed e.g. in AGNs and 
miniquasars in our Galaxy, e.g. SS433, but it is hard to imagine a
model of collapsar with beaming factor, e.g. $\sim 10^{-2}- 1-^{-3}$. 
As Blandford  said\cite{Bland}:\\
``Are GRBs beamed? ...\\
the argument that the bursts must be beamed, otherwise they would have
energies in excess of stellar rest mass, reminds me of a similar
argument in favor of them being local!''

Although general principles allow collapsars as GRB engines with 
required energy release up to ${\cal E}_{GRB}\sim ~10^{54}$~erg, there
are no models which practically realize this possibility. We shall
describe shortly four recent collapsar models for GRB engines. 

The most elaborated model is a binary neutron star merger 
NS-NS, or a similar binary system from neutron star and black hole, 
NS-BH. This  was first suggested in ref.\cite{EiSch}, for the recent
calculations and references see \cite{Janka}. Merging of two NSs , or 
NS and BH results in collapse with emission of neutrinos. The latter 
annihilate $\nu+\bar{\nu}\to e^++e^-$, producing relativistically
expanding fireball. Numerical simulations \cite{Janka} give the total
energy of neutrinos in the burst $\sim 3\times 10^{52}$~erg, but
annihilation efficiency is very low (1-3\%). Apart from small total 
energy ${\cal E}_{GRB} \sim 10^{51}$erg this model predicts very short
burst duration $\tau \leq 0.1$~s, which corresponds to a small
fraction of the observed GRBs.    

The second model\cite{Woo}, ``failed SN'' starts with the collapse of 
single rotating Wolf-Rayet star. As the core of this star collapses,
it accrets the gas from the mantle. Neutrinos interact with accretion
disc, producing a fireball. The total energy transferred to fireball 
is estimated as ${\cal E}_{GRB} \sim 10^{51}$~erg. One cannot expect
large beaming in this model. Duration of the bursts is rather short,
less than 10~s. The model predicts too large baryon contamination,
in contradiction with hydrodynamic part of the model. This model 
is based on the estimates and should be called scenario.

The two models described above, use neutrino radiation which has low
efficiency of conversion into fireball energy. An interesting collapse
scenario (hypernova) without involving neutrinos was suggested in 
Ref.\cite{Pacz}.
A massive black hole with mass $M_{bh} \sim 10 M_{\odot}$ is produced
in the end of evolution of single star. It has superstrong magnetic field 
$B \sim 10^{15}$~G and fast rotation corresponding to rotational
energy $E_{rot} \sim 5\times 10^{54}$~erg. Magnetic field by its
pressure expels the outer shell. A small fraction of it moving in the
region with decreasing density is accelerated to relativistic velocities
by the Colgate mechanism. The hypernova  scenario does not need narrow beaming 
and predicts large GRB duration. However, it predicts too large
baryonic contamination. This scenario involves a possible sequence
of physical phenomena taken with extreme values of parameters.  
It is hard to perform numerical calculations for such scenario. 

The weakness of hypernova scenario (large baryon contamination) is
evaded in Supranova scenario \cite{StVie}. The massive neutron star is
stabilised by rotation. The loss of angular momentum results in collapse
like in hypernova scenario, and a fireball is produced by expanding
magnetic field. In this scenario the baryon contamination is low,
it is provided by swept up baryons from interstellar region. 

In conclusion, the astrophysical models for GRB engine, which allow
numerical calculations, predict too low energy release and need narrow beaming,
which does not naturally exist in these models. In some scenarios,
most notably hypernova, there could be a set of physical phenomena, which when
taken with extreme parameters can provide the energy release and 
durations needed for explanation of observed GRBs. However, as it stands, 
these scenarios do not give numerical predictions and, strictly
speaking, cannot be called models. 
\\*[2mm]
{\bf 5.1 GRBs from superconducting strings}
\\*[2mm]
Can the problem of GRB engine with its tremendous energy release,
probably beamed, be solved with a help of a new physics?  

Such a solution, utilising the cusps of superconducting strings, was
first suggested in Refs.\cite{BPS} and recently revived in Ref.\cite{BHV}. 

Cosmic strings are linear defects that could be formed at a symmetry
breaking phase transition in the early universe \cite{Book}. In the
space they exist in the form of endless strings and closed loops. Strings
predicted in most grand unified models respond to external
electromagnetic fields as thin superconducting wires \cite{Witten}.
As they move through cosmic magnetic fields, such strings develop
electric currents. Oscillating loops of
superconducting string emit short bursts of highly beamed
electromagnetic radiation through small string segments, centered at
peculiar points on a string, cusps, where velocity reaches speed of 
light \cite{VV,SPG}.

The beam of low-frequency e-m radiation propagating in plasma produces
a beam of accelerated particles.  For an e-m wave in vacuum,
a test particle would be
accelerated to a very large Lorentz
factor. But the maximum Lorentz factor of the {\em beam}  is saturated
at the value $\gamma_b$, when the energy of the
beam reaches the energy of the original e-m pulse:
$N_b m \gamma_b \sim {\cal E}_{em}$. This results in the Lorentz factor
of the beam of order $10^2 - 10^4$ like in ordinary fireball. 
The beam of accelerated particles pushes the
gas with the frozen magnetic field ahead of it, producing an external
shock in surrounding plasma  and  a reverse  shock
in the beam material, as in the case of ordinary fireball. Therefore 
the hydrodynamic development of a fireball is essentially the same 
as in case of astrophysical GRB engine.  

In contrast to astrophysical models, a cusp produces pulse of e-m 
radiation with energy and beaming fixed by parameters of cosmic
strings and ambient gas. In fact, the cusp model is very rigid. 
In a simplified version \cite{BHV} there is only one free parameter,
the string scale of symmetry breaking $\eta \sim 10^{14}$~GeV, and 
two physical quantities which characterise gas in filaments and sheets,  
where most of GRBs are originated. Formally there are three such 
quantities: 
magnetic field, parametrised as $B=B_{-7}10^{-7}$~G, redshift of 
its origin $z_m$ and gas density $n_g$. But  dependence
on gas density is extremely weak. It influences only hydrodynamic
flow through Lorentz factor of contact discontinuity surface , 
$\gamma_{CD}$, with $\gamma_{CD} \propto n_g^{-1/8}$. 

The only genuine free parameter, $\eta$, or equivalently dimensionless
parameter $\alpha=k_g G \eta^2$, where $G$ is a gravitational constant
and $k_g \sim 50$ is a numerical coefficient, determines the space 
properties of the loops: a typical length of a loop at epoch $t$, 
$l \sim \alpha t$ and the number density of loops  
\begin{equation}
n_l(t) \sim \alpha^{-1}t^{-3}.  
\end{equation}

A string segment near the cusp moves with Lorentz factor $\gamma$ and
radiates e-m pulse within a cone with opening angle $\theta \sim 1/\gamma$.
The energy radiated per unit of solid angle in the direction $\theta$
is given by \cite{VV}
\begin{equation}
d{\cal E}_{em}/d\Omega \sim k_{em}J_0^2 \alpha t /\theta^3,
\label{Emax}
\end{equation}
where $J_0$ is initial current induced by external magnetic field in a loop.

The fluence, defined as the total energy
per unit area of the detector, is 
\begin{equation}
S\sim (1+z)(d{\cal E}_{em}/d\Omega) d_L^{-2}(z),
\label{S}
\end{equation}
where $d_L(z)=3t_0(1+z)^{1/2}[(1+z)^{1/2}-1]$ is the luminosity
distance.

After simple calculations \cite{BHV} one obtains the GRB rate as a
function of fluence. For relatively small fluences,
\begin{equation}
{\dot N_{GRB}(>S)} \approx 3\cdot 10^2 S_{-8}^{-2/3}B_{-7}^{4/3}~yr^{-1},
\label{rate}
\end{equation}
while for large fluences $\dot N_{GRB}(>S) \propto S^{-3/2}$.
Both absolute value of  $\dot N_{GRB}(>S)$ and its dependence on $S$
agree with observations.
The duration of the cusp event as seen by a distant observer is \cite{BPS}
\begin{equation}
\tau_c\sim (1+z)(\alpha t/2)\gamma^{-3}
\sim (\alpha t_0/2) (1+z)^{-1/2}\theta^{3}.
\label{tau}
\end{equation}
The duration of the cusp event coincides with the duration of GRB, 
found as duration of fireball emission \cite{BHV}. 
The duration of GRBs originating at redshift $z$ and having fluence
$S$ can be readily calculated as
\begin{equation}
\tau_{GRB} \approx 200 \frac{\alpha_{-8}^4 B_{-7}^2}{S_{-8}}
(1+z)^{-1}(\sqrt{1+z}-1)^{-2}~s
\label{duration}
\end{equation}
Analysis of Eq.(\ref{duration}) shows \cite{BHV} that it correctly
describes the range of observed GRB durations. 

Therefore, simplified one-parameter cusp model describes correctly 
the total energies of GRBs (or fluences $S$), the GRB rates and their
dependence on $S$, and GRB durations (absolute values and a range).

The signatures of this model are simultaneous powerful 
bursts of gravitational radiation \cite{BHV,DaVi} from a cusp and 
repeaters for GRBs of very short durations.

The cusp model meets basically one difficulty: it predicts too low
GRB rate from galaxies. This discrepancy could be eliminated if the
model strongly underestimates the capture rate of string loops by
galaxies.  For example, if $\alpha\gg k_g G\eta^2$, then the loops are
non-relativistic and may be effectively captured by galaxies.
Another possibility is that the cusp model could 
describes some subclass of the sources not associated with galaxies.
Such a subclass could include the short-duration GRBs for which host
galaxies are not found, or  another subclass of
no-host GRBs. 

\section{Conclusions}

Astronomy and astrophysics were in the past and remain now a source of 
fundamental discoveries in physics. What phenomena can we suspect now
as the challengers for such discoveries? 

Most probably UHECR is a challenger number one. Presence of detected
particles with energies above the GZK cutoff is reliably established. 
Astrophysical solution to UHECR problem is disfavoured,
though not excluded. UHE heavy nuclei (iron) accelerated in our
Galaxy, local (10 -30 Mpc) enhancement of UHECR sources and exotic  
one-source models remain disfavoured but not rigorously excluded
possibilities. One might  expect surprises here, but each of them will be
a small revolution in a special field. There are many
elementary-particle solutions to UHECR problem. All of them seem
exotic to non-specialists, but eventually many of them not. For
example, idea of UHECR from Superheavy DM is based only on known   
theoretical physics. Light gluino as a carrier of UHE signal is also 
based on reliable physics and may be saved by re-examination 
or re-interpretation of accelerator data. Only future experiments can 
can solve UHECR problem. The observations of UHECR in the
southern hemisphere (the Auger detector), measurement of longitudinal
profiles of EAS in the atmosphere (High-Res) and search for the 
particles  with energies $1\times 10^{21}$~eV and higher (Auger, Telescope
Array and space detectors) may result in fundamental discovery 
in physics.\\*[2mm]
GRB engines with their tremendous energy output, most probably beamed, 
are new type of the sources even in case they are collapsars. 
Further observational evidences for their origin ( more precise 
positions in the 
galaxies, no-host GRBs etc) will help to establish the nature of 
these objects, though they are well hidden in the debris of the
bursts. Search for gravitational bursts and very short GRBs can bring 
evidence for non-astrophysical origin of at least part of 
GRBs. 
\\*[2mm]
I think TeV gamma-ray crisis will be peacefully solved by
reconsideration of observed diffuse flux of IR radiation. It will
result then in a better constraint on the Lorentz Invariance violation, not 
in its discovery. 

\section{Acknowledgements}

It is my great pleasure to thank Milla Baldo Ceolin for organising
most exciting conference and for magnificent hospitality.\\   
This work has been performed as part of INTAS project No. 99-1065.


\begin{thebibliography}{99}
\bibitem{SN1987}
K.Hirata et al., Phys. Rev. Lett. {\bf 58} (1987) 1490,\\
R.M.Bionta et al., Phys. Rev. Lett. {\bf 58} (1987) 1494,\\
E.N.Alexeev et al., Sov. Astron. Lett. {\bf 14} (1988) 41.

\bibitem{ImNa}
V.S.Imshennik and D.K.Nadyozhin, Sov. Astr. Lett. {\bf 3} (1977) 353.

\bibitem{string}
V.A.Kostelecki and S.Samuel, Phys. Rev. {\bf D39} (1989)  683.

\bibitem{Dvali}
G.Dvali and M.Shifman, Phys. Rept. {\bf 320} (1999) 107.

\bibitem{Ellis}
J.Ellis, N.E.Mavromatos and D.V.Nanopolos, 4th International
Symposium On Sources And Detection Of Dark Matter In The Universe (DM 2000), 
Marina del Rey, California, 20-23 Feb 2000, gr-qc/0005100.

\bibitem{CoKo97}
Don Colladay and V.A.Kostelecky, Phys.Rev. {\bf D55}, (1997) 6760.

\bibitem{CoGl}
S.Colemam and S.L. Glashow, Phys.Lett.{\bf B405} (1997) 249;\\
S.Coleman and S.L. Glashow, Phys.Rev. {\bf D59} (1999) 116008.


\bibitem{KG}
V.B.Berestetskii, E.M.Lifshitz and L.P.Pitaevskii, {\it Quantum 
Electrodynamics}, Pergamon Press, 1980.

\bibitem{GZK}
K.Greisen, Phys.Rev.Lett., {\bf 16} (1966) 748;\\
G.T.Zatsepin and V.A.Kuzmin, Pisma Zh. Experim. Theor. Phys., {\bf 4}
(1966) 114.


\bibitem{Kirzh}
D.A.Kirzhnitz and V.A.Chechin, Pisma in ZhETP, {\bf 14} (1972) 261;
D.A.Kirzhnitz and V.A.Chechin, Sov. Journal of Nucl. Phys., {\bf 15} (1972)
1051

\bibitem{Gonz}
L.Gonzales-Mestres, in {\it Proc. 25th Inter. Cosmic Ray Conf.},
Durban, {\bf 6} (1997) 109;\\
L.Gonzales-Mestres, in {\it Proc. 26th Inter. Cosmic Ray
Conf.}, Salt Lake City, {\bf 1} (1999) 179.

\bibitem{Glash}
S.L.Glashow, Nucl.Phys.B (Proc. Suppl.) {\bf 70} (1999) 180. 


\bibitem{Grillo}
R.Aloisio, P.Blasi, P.L.Ghia and A.F.Grillo, Phys. Rev. {\bf D62} (2000) 053010.

\bibitem{ElNa}
J.Ellis et al, Astrophys.J. {\bf 535} (2000) 139.


\bibitem{DIRBE}
M.G.Hauser et al, Ap.J. {\bf 508} (1998) 25.


\bibitem{FIRAS}
D.J.Fixsen et al, Ap.J. {\bf 508} (1998) 123


\bibitem{ISOCAM}
A.Biviano et al, astro-ph/9910314


\bibitem{Steck}
M.A.Malkan and F.W.Stecker, astro-ph/0009500,\\
F.W.Stecker, IAU Symposium (eds M.Harwit and G.Hauser) {\bf 204} (2000),
astro-ph/0010015

\bibitem{Prim}
J.R.Primack, Astrop. Phys., {\bf 11} (1999) 93,\\
D.MacMinn and J.R.Primack, Sp.Sci.Rev. {\bf 75} (1996) 413.

\bibitem{Proth}
R.J.Protheroe and N.Meyer, Phys. Lett. {\bf B 493} (2000) 1.

\bibitem{Ahar}
J.Guy, C.Renault, F.A.Aharonian, M. Rivoal and J.-P.Tavernet, 
Astron.Astroph. {\bf 359} (2000)419.

\bibitem{BGZ}
V.S.Berezinsky,S.I.Grigorieva and G.T.Zatsepin, 
Proc. 14th Int.Cosm.Ray Conf.(Munick) {\bf 2} (1975) 711;\\
Izv.Acad.Nauk USSR (ser. phys) {\bf 40} (1976) 524.

\bibitem{Be1970}
V.S.Berezinsky, Soviet Phys.Nucl.Phys., {\bf 11} (1970) 399.
\bibitem{NaWa}
M.Nagano and A.A.Watson, Rev. of Mod. Phys. {\bf 72} (2000) 689.


\bibitem{Hillas}
A.M.Hillas et al, 12th Int. Cosm.Ray Conf. {\bf 3} (1971) 1001.


\bibitem{FE}
D.J.Bird et al, Ap.J. {\bf 424} (1994) 491.

\bibitem{AGASA}
S.Yoshida et al, Astrop. Phys. {\bf 3} (1995) 105. 

\bibitem{book}
V.S. Berezinskii, S.V.Bulanov, V.A.Dogiel, V.L.Ginzburg, and
V.S.Ptuskin, {\it Astrophysics of Cosmic Rays}, North-Holland, 1990.

\bibitem{BGG}
V.Berezinsky, A.Gazizov and S.Grigorieva, to be published in Proc. of
ICRC 2001.

\bibitem{SS}
S.T.Scully and F.W.Stecker, astro-ph/0006112.

\bibitem{BGLS}
V.S.Berezinsky and S.I.Grigorieva, Proc. of 16th Int. Cosmic Ray
Conf. (Kyoto), {\bf 2}, (1979) 81.

\bibitem{BlOl}
M.Blanton, P.Blasi and A.Olinto, Astrop. Phys. {\bf 15} (2001) 275.

\bibitem{Wo}
J.Wdowczyk and A.Wolfendale, Nature {\bf 281} (1980) 356;\\
M.Giler,J.Wdowczyk and A.Wolfendale, J.Phys.G {\bf 6} (1980) 1561.

\bibitem{Bier}
E-J Ahn, G.Medino-Tanco, P.L.Biermann and T.Stanev,
Nucl.Phys. (Proc. Suppl.) {\bf 87} (2000) 417.

\bibitem{BGD}
V.S.Berezinsky, S.I.Grigorieva and V.A.Dogiel, Astron.Astroph., 
{\bf232} (1990) 582.

\bibitem{Sy}
S.I.Syrovatsky, Comments Ap.Phys. {\bf 3} (1969) 155.

\bibitem{Chr}
G.V.Kulikov, Yu.A.Fomin and G.B.Khristiansen, ZhETP Lett.,{\bf 11} 
(1969) 374.

\bibitem{Gill}
M.Giller and M.Zielinska, Proc. 25th Int.Cosm.Ray.Conf. (Durban)
{\bf 4} (1997) 469,

\bibitem{Pt}
V.N.Zirakashvili et al, Astro. Lett. {\bf 24} (1998) 139.

\bibitem{BGM} 
V.Berezinsky, S,Grigorieva and M.Marchesini, im preparation.

\bibitem{COSMAG}
V.S.Berezinsky et al, in {\it Astrophysical Aspects of Most Energetic
Cosmic Rays}, ed. M.Nagano and F.Takahara, World Scientific, (1991) 134.

\bibitem{SS}
D.D.Sokoloff and A.M.Shukurov, Nature {\bf 347} (1990) 51.

\bibitem{Be99}
V.Berezinsky, Nucl.Phys.B (Proc.Suppl) {\bf 87} (2000) 387.

\bibitem{Bland}
R.D.Blandford, astro-ph/0001498.

\bibitem{EiSch}
D.Eichler, M.Livio, T.Piran and D.N.Schramm, Nature {\bf 340} (1989) 126.

\bibitem{Janka}
H-Th.Janka et al, Ap.J. {\bf 527} (1999) L39; and astro-ph/0101357.

\bibitem{Woo}
S.E.Woosley, Ap.J. {\bf 405} (1993) 273; Ap.J. {\bf 527} (1999) 

\bibitem{Pacz}
B.Paczinski, Ap.J. {\bf 494} (1998) L45, astro-ph/9706232.

\bibitem{StVie}
L.Stella and M.Vietri, Ap.J. {\bf 507} (1998) L45; Ap.J.{\bf 527} (1999) L43.
 
\bibitem{BPS}
A.Babul, B.Paczynski and D.N.Spergel, Ap.J.Lett. {\bf 316}, L49 (1987),\\
B.Paczynski, Ap.J. {\bf 335}, 525 (1988).

\bibitem{BHV}
V.Berezinsky, B.Hnatyk and A.Vilenkin, astro-ph/0102366.

\bibitem{Book}
A. Vilenkin and E.P.S. Shellard, {\it Cosmic strings
and other topological defects}, Cambridge University Press, Cambridge,
1994.

\bibitem{Witten} E. Witten, Nucl. Phys. {\bf B249} (1985) 557.

\bibitem{VV}
A.Vilenkin and T.Vachaspati, Phys.Rev.Lett. {\bf 58} (1987) 1041.

\bibitem{SPG} D.N.Spergel, T.Piran and J.Goodman, Nucl.Phys. {\bf B291}, 
(1987) 847.

\bibitem{DaVi}
T.Damour and A.Vilenkin, Phys.Rev.Lett. {\bf 85}, 3761 (2000)

\end{thebibliography}
\end{document}